\documentstyle[psfig]{l-aa}

\begin{document}

\thesaurus{ 06(08.12.1; 08.12.2; 13.25.5) }

\title{ Search for X-ray emission from bona-fide and candidate brown dwarfs }

\author{ R. Neuh\"auser\inst{1}, C. Brice\~no\inst{2}, 
F. Comer\'on\inst{3}, T. Hearty\inst{1}, 
E.L. Mart\'\i n\inst{4}, J.H.M.M. Schmitt\inst{5}, 
B. Stelzer\inst{1}, R. Supper\inst{1},
W. Voges\inst{1}, \and H. Zinnecker\inst{6} }

\offprints{R. Neuh\"auser, rne@mpe.mpg.de \\
co-authors are listed in alphabetical order }

\institute{MPI f\"ur extraterrestrische Physik, Giessenbachstra\ss e 1, D-85740 Garching, Germany
\and Yale University, Department of Physics, New Haven, CT 06520-8121, USA
\and European Southern Observatory, Karl-Schwarzschild-Stra\ss e 2, D-85748 Garching, Germany
\and Astronomy Department, University of California at Berkeley, Berkeley, CA 94720, USA
\and Universit\"at Hamburg, Sternwarte, Gojensbergweg 112, D-21029 Hamburg, Germany 
\and Astrophysikalisches Institut, An der Sternwarte 16, D-14482 Potsdam, Germany
}

\date {Received 24 Sep 1998; accepted 10 Dec 1998 }

\maketitle

\markboth{Neuh\"auser et al.: Search for X-ray emission from brown dwarfs }{}

\begin{abstract}

Following the recent classification of the X-ray detected object 
V410 x-ray 3 with a young brown dwarf candidate (Brice\~no et al. 1998)
and the identification of an X-ray source in Chamaeleon as
young bona-fide brown dwarf (Neuh\"auser \& Comer\'on 1998),
we investigate all ROSAT All-Sky Survey and archived ROSAT 
PSPC and HRI pointed observations with bona-fide or candidate brown dwarfs 
in the field of view with exposure times ranging from 0.13 to 221~ks, 
including dedicated 64~ks and 42~ks deep ROSAT HRI pointed observations 
on the low-mass star BRI~0021$-$0214 and the brown dwarf Calar 3, respectively.
Out of 26 bona-fide brown dwarfs, one is newly detected in X-rays,
namely $\rho$ Oph GY~202.
Also, four out of 57 brown dwarf candidates studied here 
are detected in X-rays, namely the young Taurus brown dwarf 
candidates MHO-4, MHO-5, V410 Anon 13, and V410 x-ray 3.
The M9.5-type star BRI~0021$-$0214 is not detected. 
In the appendix, we also present catalogued, but as yet unnoticed 
B- and R-band data for some of the objects studied here.


\end{abstract}

\section {Introduction}

Objects which are unable to sustain stable nuclear fusion of hydrogen, 
but which can burn deuterium until they are $\sim 10^{7}$ yrs old, 
are called brown dwarfs; see Kulkarni (1997) for a recent review.
They continue to contract until electron degeneracy halts further contraction.
Depending on metallicity and model assumptions made in for calculating 
theoretical evolutionary tracks, the limiting mass between normal stars 
and brown dwarfs is $\sim 0.075$ to $0.08~M_{\odot}$ (Burrows et al. 1995, 1997, 
D'Antona \& Mazzitelli 1994, 1997, Allard et al. 1997, Baraffe et al. 1998).
Objects with masses below $\sim 0.01~M_{\odot}$ cannot even burn deuterium and 
are called planets. One can also distinguish brown dwarfs from planets 
by the formation mechanism, namely objects formed in circumstellar accretion 
disks would be called planets, while those which formed by fragmentation 
of a protostellar cloud would be called brown dwarfs; 
however, we prefer to link the distinction with the
physics going on inside the objects rather than with its formation,
so that we regard all objects between $\sim 0.01$ and $0.08~M_{\odot}$ 
as brown dwarfs, regardless of how they formed.
One can identify an object as a brown dwarf either using the Lithium test
(an object with primordial Lithium abundance either is very young or a 
brown dwarf or both, Rebolo et al. 1992) or by finding an object 
below the stellar limit in an H-R
or color-magnitude
diagram (for this, one needs to know the distance, eg., 
due to a bright stellar companion or confirmed membership of a cluster).


\begin{table}

\begin{tabular}{lclcl} 
\multicolumn{5}{c}{ \bf Table 1: Brown dwarfs (except Cha I). } \\ \hline
Designation & \hspace{-1cm} $\log$ $L_{bol} / L_{\odot}$ & area & \hspace{-.5cm} dist. [pc] & ref. \\ \hline

Roque 4    & $-3.35$ & Pleiades & \hspace{-.5cm} 125 & 1   \\ 
MHObd3     & $-3.03$ & Pleiades & \hspace{-.5cm} 125 & 2   \\ 
Roque 5    & $-3.45$ & Pleiades & \hspace{-.5cm} 125 & 3   \\ 
Roque 13   & $-3.00$ & Pleiades & \hspace{-.5cm} 125 & 1   \\ 
Roque 11   & $-3.15$ & Pleiades & \hspace{-.5cm} 125 & 1   \\ 
Teide 1    & $-3.18$ & Pleiades & \hspace{-.5cm} 125 & 4,5 \\ 
Roque 17   & $-2.83$ & Pleiades & \hspace{-.5cm} 125 & 1   \\ 
Roque 16   & $-2.89$ & Pleiades & \hspace{-.5cm} 125 &1,2,6\\ 
PPl 15     & $-2.80$ & Pleiades & \hspace{-.5cm} 125 & 7   \\ 
Roque 12   & $-3.14$ & Pleiades & \hspace{-.5cm} 125 & 3   \\ 
Roque 25   & $-3.90$ & Pleiades & \hspace{-.5cm} 125 & 3   \\ 
Calar 3    & $-3.11$ & Pleiades & \hspace{-.5cm} 125 & 4,6 \\ 
Teide 2    & $-2.90$ & Pleiades & \hspace{-.5cm} 125 & 8   \\ 
CFHT-PL-18 & $-3.07$ & Pleiades & \hspace{-.5cm} 125 & 9,10\\ 
CFHT-PL-12 & $-2.82$ & Pleiades & \hspace{-.5cm} 125 & 6,9 \\ 
CFHT-PL-15 & $-3.16$ & Pleiades & \hspace{-.5cm} 125 & 6,9 \\ 
LP~944-20  & $-3.84$ & field & \hspace{-.5cm} 5 & 11,12 \\ 
DenisJ1228$-$1547 & $-4.30$ & field & \hspace{-.5cm} 13 & 13,14 \\ 
Kelu 1     &         & field & \hspace{-.5cm} $\ge 12$ & 15 \\ 
CRBR 14    & $-1.52$ & $\rho$ Oph & \hspace{-.5cm} 160 & 16,17 \\ 
GY 10      & $-1.38$ & $\rho$ Oph & \hspace{-.5cm} 160 & 16,17 \\ 
GY 11 (*)  & $-2.63$ & $\rho$ Oph & \hspace{-.5cm} 160 & 16,17 \\ 
GY 64      & $-2.01$ & $\rho$ Oph & \hspace{-.5cm} 160 & 16,17 \\ 
GY 141     & $-2.36$ & $\rho$ Oph & \hspace{-.5cm} 160 & 16,18 \\ 
GY 202     & $-1.84$ & $\rho$ Oph & \hspace{-.5cm} 160 & 16,17 \\ 
GY 310     & $-1.20$ & $\rho$ Oph & \hspace{-.5cm} 160 & 16,17 \\ \hline 

\end{tabular}

\vspace{-.3cm} 

{\small
Remark: (*) $L_{bol}$ is uncertain, because this object seems to be variable in
the infrared (see Comer\'on et al. 1993, 1998a, Wilking et al. 1999). \\
References: (1) Zapatero-Osorio et al. 1997b, (2) Stauffer et al. 1998a,
(3) Mart\'\i n et al. 1998b, (4) Rebolo et al. 1995, (5) Rebolo et al. 1996, 
(6) Zapatero-Osorio et al. 1997a, (7) Basri et al. 1996, 
(8) Mart\'\i n et al. 1998a, (9) Bouvier et al. 1998, 
(10) Mart\'\i n et al. 1998c, (11) Kirkpatrick et al. 1997, 
(12) Tinney 1998, (13) Delfosse et al. 1997, (14) Mart\'\i n et al. 1997, 
(15) Ruiz et al. 1997, (16) Comer\'on et al. 1993, 
(17) Wilking et al. 1999, (18) Luhman et al. 1997
}

\end{table}



\begin{table}

\begin{tabular}{lclcl} 
\multicolumn{5}{c}{ \bf Table 2: Brown dwarf candidates (except Cha I). } \\ \hline
Designation & \hspace{-1cm} $\log$ $L_{bol} / L_{\odot}$ & area & \hspace{-.5cm} dist. [pc] & ref. \\ \hline

PC~0025$+$0447 & $-3.74$ & field & \hspace{-.5cm} 60 & 18 \\ 
296~A       & $-2.88$ & field & \hspace{-.5cm} 45 & 19 \\ 
DenisJ0205$-$1159 & $-4.00$ & field & \hspace{-.5cm} 18 & 12 \\ 
AP 270     & $-2.66$ & $\alpha$ Per & \hspace{-.5cm} 170 & 20,21 \\ 
CFHT-PL-8   & $-2.87$ & Pleiades & \hspace{-.5cm} 125 & 8 \\ 
CFHT-PL-17  & $-3.17$ & Pleiades & \hspace{-.5cm} 125 & 8 \\ 
Roque 7     & $-3.31$ & Pleiades & \hspace{-.5cm} 125 & 8 \\ 
CFHT-PL-20  & $-3.24$ & Pleiades & \hspace{-.5cm} 125 & 8 \\ 
CFHT-PL-16  & $-3.01$ & Pleiades & \hspace{-.5cm} 125 & 8 \\ 
MHObd4      & $-3.05$ & Pleiades & \hspace{-.5cm} 125 & 22 \\ 
MHObd1      & $-2.90$ & Pleiades & \hspace{-.5cm} 125 & 22 \\ 
CFHT-PL-19  & $-3.16$ & Pleiades & \hspace{-.5cm} 125 & 8 \\ 
Roque 15    & $-2.86$ & Pleiades & \hspace{-.5cm} 125 & 1 \\ 
Roque 14    & $-3.00$ & Pleiades & \hspace{-.5cm} 125 & 1 \\ 
NPL 37      & $-3.22$ & Pleiades & \hspace{-.5cm} 125 &23,24\\ 
MHObd5      & $-2.90$ & Pleiades & \hspace{-.5cm} 125 & 22 \\ 
NPL 38      & $-3.24$ & Pleiades & \hspace{-.5cm} 125 &23,24\\ 
PIZ~1       & $-3.39$ & Pleiades & \hspace{-.5cm} 125 & 25 \\ 
NPL 36      & $-3.14$ & Pleiades & \hspace{-.5cm} 125 &23,24\\ 
CFHT-PL-5   & $-2.71$ & Pleiades & \hspace{-.5cm} 125 & 8 \\ 
NPL 40      & $-3.50$ & Pleiades & \hspace{-.5cm} 125 &23,24\\ 
MHObd6      & $-2.85$ & Pleiades & \hspace{-.5cm} 125 & 22 \\ 
CFHT-PL-1   & $-2.43$ & Pleiades & \hspace{-.5cm} 125 & 8 \\ 
CFHT-PL-7   & $-2.86$ & Pleiades & \hspace{-.5cm} 125 & 8 \\ 
CFHT-PL-6   & $-2.56$ & Pleiades & \hspace{-.5cm} 125 & 8 \\ 
CFHT-PL-23  & $-3.27$ & Pleiades & \hspace{-.5cm} 125 & 8 \\ 
CFHT-PL-2   & $-2.58$ & Pleiades & \hspace{-.5cm} 125 & 8 \\ 
HHJ~22      & $-2.61$ & Pleiades & \hspace{-.5cm} 125 & 8,26 \\ 
CFHT-PL-4   & $-2.65$ & Pleiades & \hspace{-.5cm} 125 & 8 \\ 
CFHT-PL-26  & $-3.63$ & Pleiades & \hspace{-.5cm} 125 & 8 \\ 
CFHT-PL-25  & $-3.42$ & Pleiades & \hspace{-.5cm} 125 & 8 \\ 
CFHT-PL-22  & $-3.39$ & Pleiades & \hspace{-.5cm} 125 & 8 \\ 
V410 x-ray 3& $-1.21$ & Taurus   & \hspace{-.5cm} 140 & 27,28 \\ 
V410 Anon 13& $-1.78$ & Taurus   & \hspace{-.5cm} 140 & 28 \\ 
Tau MHO-4   & $-1.37$ & Taurus   & \hspace{-.5cm} 140 & 28 \\ 
Tau MHO-5   & $-1.53$ & Taurus   & \hspace{-.5cm} 140 & 28 \\ 
RPr 1       & $-3.60$ & Praesepe & \hspace{-.5cm} 177 & 29 \\ 
CRBR 15     & $-1.32$ & $\rho$ Oph & \hspace{-.5cm} 160 & 15,16  \\ 
GY 5        & $-1.14$ & $\rho$ Oph & \hspace{-.5cm} 160 & 16 \\
CRBR 28     & $-1.29$&$\rho$ Oph & \hspace{-.5cm} 160 & 15,30 \\ 
CRBR 33     & $-1.52$&$\rho$ Oph & \hspace{-.5cm} 160 & 15,30 \\ 
GY 31       & $-0.09$ & $\rho$ Oph & \hspace{-.5cm} 160 & 16 \\
GY 37       & $-1.56$ & $\rho$ Oph & \hspace{-.5cm} 160 & 16 \\
GY 59       & $-1.23$ & $\rho$ Oph & \hspace{-.5cm} 160 & 16 \\
GY 84       & $-0.93$ & $\rho$ Oph & \hspace{-.5cm} 160 & 16 \\
GY 107      & $-0.81$ & $\rho$ Oph & \hspace{-.5cm} 160 & 16 \\
GY 163      & $-0.69$ & $\rho$ Oph & \hspace{-.5cm} 160 & 16 \\
GY 218      & $-2.06$ & $\rho$ Oph & \hspace{-.5cm} 160 & 15,30 \\ 
GY 326      & $-0.64$ & $\rho$ Oph & \hspace{-.5cm} 160 & 16 \\
B185815.3$-$370435 & $-2.00$ & CrA & \hspace{-.5cm} 130 & 31 \\ 
B185831.1$-$370456 & $-2.60$ & CrA & \hspace{-.5cm} 130 & 31 \\ 
B185839.6$-$365823 & $-3.10$ & CrA & \hspace{-.5cm} 130 & 31 \\ 
B185840.4$-$370433 & $-2.20$ & CrA & \hspace{-.5cm} 130 & 31 \\ 
B185853.3$-$370328 & $-2.20$ & CrA & \hspace{-.5cm} 130 & 31 \\ 
D04               & $-4.23$ & field & \hspace{-.5cm} 48 & 32 \\
D07               & $-4.53$ & field & \hspace{-.5cm} 43 & 32 \\ \hline 
BRI~0021$-$0214   & $-3.50$ & field star& \hspace{-.5cm} 12 & 33 \\ 
\end{tabular}

\end{table}

\begin{table}

\begin{tabular}{lclcl} 
\multicolumn{5}{c}{ {\bf Table 2: Brown dwarf candidates} (cont.) } \\ \hline
\end{tabular}

{\small
References: (1) to (17) as in Table 1,
(18) Schneider et al. 1991, (19) Thackrah et al. 1997, 
(20) Prosser 1994, (21) Basri \& Mart\'\i n 1998a,
(22) Stauffer et al. 1998b, (23) Festin 1998a, (24) Festin 1998b,
(25) Cossburn et al. 1997, (26) Hambly et al. 1993, 
(27) Luhman et al. 1998, (28) Brice\~no et al. 1998,
(29) Magazz\`u et al. 1998,
(30) Comer\'on et al. 1998a, (31) Wilking et al. 1997,
(32) Hawkins et al. 1998, (33) Tinney et al. 1995.
}

\end{table}


Because brown dwarfs have no stable nuclear energy source and 
derive most of their luminosity from gravitational contraction, 
they cool down and become less luminous as they age
(Burrows et al. 1995, Brandner et al. 1997, Malkov et al. 1998). 
Hence, several search programs focused on zero-age main-sequence
clusters and pre-main sequence associations.
Eg., several brown dwarfs were discovered in the $\sim 10^{8}$ yr 
old Pleiades cluster (see Tables 1 and 2 for references).
Brice\~no et al. (1998) could identify four brown dwarf candidates
in the Taurus star forming region, one of which is V410 x-ray 3, 
a faint X-ray source detected by Strom \& Strom (1994)
and studied in detail also by Luhman et al. (1998).
Several bona-fide and candidate brown dwarfs were found 
in $\rho$ Oph (Rieke \& Rieke 1990, Comer\'on et al. 1993, 1998a,
Luhman et al. 1997, Wilking et al. 1999) 
and the Cha I dark cloud (Comer\'on et al. 1999,
henceforth CRN99; Neuh\"auser \& Comer\'on 1998, NC98). 

Here, we investigate on a statistically meaningful sample whether 
or not brown dwarfs emit X-rays. This sample consists of objects with
largely differing properties, such as age, luminosity, temperature, 
and cluster membership. Thus, our study has the potential of exploring 
how X-ray properties are affected by all those factors.

In Sect. 2, we present the motivation 
for our study by elaborating on why brown dwarfs might be X-ray sources.
In Sect. 3, we list all bona-fide and candidate brown dwarfs        
published so far (Tables 1 and 2),
explain our X-ray data reduction procedures,
list the ROSAT pointed observations with bona-fide or candidate
brown dwarfs in the field of view (Table 3), and present the X-ray 
data for brown dwarfs and candidates (Tables 4 to 6) including the 
X-ray light curves for the two detected brown dwarf candidates 
with sufficient counts for a meaningful timing analysis.
After comparing our results with those obtained by NC98 and CRN99 
in Cha I (Sect. 4), we conclude with a brief discussion in Sect. 5.

\section {Motivation}

Any X-ray detection or low upper limit of a bona-fide 
brown dwarf would improve our currently incomplete understanding 
of brown dwarfs, and in particular it would help answering
the question whether brown dwarfs support 
some mechanism capable of heating coronae. We recall that T~Tauri stars 
and also optically invisible infrared Class I objects (protostars) display 
X-ray activity, despite the lack of a stable nuclear energy source 
(see Neuh\"auser 1997 for a review). 
Gravitationally contracting low-mass pre-main sequence objects,
at least those with spectral type M, are fully convective, 
and in some sense similar to young brown dwarfs 
which - by definition - never reach the main sequence, 
so that we may suspect young brown dwarfs 
to display X-ray emission via a similar mechanism.

While it is not clear whether brown dwarfs can drive and sustain 
a dynamo capable of heating coronae similar to low-mass stars, 
it is encouraging that 
some very old late-type stars are also detected in X-rays:
vB~8 with spectral type M7 and a mass of $\sim 0.08~M_{\odot}$ 
is clearly detected as an X-ray source (Fleming et al. 1993, 
Giampapa et al. 1996). Fleming et al. (in preparation)
detected an X-ray flare of the M8-type star vB~10, 
which is not detected before and after the flare.


\begin{table}

\begin{tabular}{rlclr} 
\multicolumn{5}{c}{ \bf Table 3: ROSAT pointings analyzed here. } \\ \hline
no. & Pointing ID & Instr. & PI & exp. [s] \\ \hline 

1  & 201077p                & PSPC & Fleming     & 9065 \\
2  & 000105p                & PSPC & MPE         & 650 \\
3  & 200008p\,-\,0\,/\,-\,2 & PSPC & Caillault & 12985 \\
4  & 200008p\,-\,-\,2       & PSPC & Caillault & 696 \\
5  & 200008p\,-\,-\,3       & PSPC & Caillault &  45 \\
6  & 200008p\,-\,-\,4       & PSPC & Caillault & 121 \\
7  & 200008p\,-\,-\,5       & PSPC & Caillault & 722 \\
8  & 200068p\,-\,0\,/\,-\,1 & PSPC & Rosner  & 39920 \\
9  & 200068p\,-\,-\,1       & PSPC & Rosner    & 1307 \\
10 & 200556p                & PSPC & Stauffer  & 22456 \\
11 & 200557p                & PSPC & Stauffer  & 27648 \\
12 & 200949p                & PSPC & Feigelson & 6098 \\
13 & 200001p\,-\,0\,/\,-\,1 & PSPC & Strom     & 30077 \\
14 & 201598p\,-\,201602p    & PSPC & Barwig    & 28973 \\
15 & 201312p                & PSPC & Zinnecker & 2800 \\
16 & 201025p                & PSPC & Walter    & 5448 \\
17 & 900353p                & PSPC & Burrows   & 7718 \\
18 & 201313p                & PSPC & Zinnecker & 4027 \\
19 & 200443p                & PSPC & Pye       & 20074 \\
20 & 600043p                & PSPC & Jones     & 53511 \\
21 & 600127p                & PSPC & Petre     & 17766 \\
22 & 800301p\,/\,-\,1       & PSPC & Loewenstein & 8917 \\
23 & 700921p\,/\,-\,1       & PSPC & Turner    & 10306 \\ 
24 & 200250p                & PSPC & Schmitt   & 1897 \\
25 & 200599p\,-\,200625p    & PSPC & Schmitt   & 30166 \\
26 & 200045p\,-\,0\,/\,-\,1 & PSPC & Montmerle & 32847 \\ 
27 & 200493p\,-\,0\,/\,-\,1 & PSPC & Walter    & 7460 \\
28 & 900002p                & PSPC & Garmire   & 7761 \\ \hline
29 & 202214h\,-\,1\,/\,-\,2 & HRI & Schmitt   & 63897 \\
30 & 400764h                & HRI & Halpern   & 17615 \\
31 & 600256h                & HRI & Kim       &  7339 \\
32 & 600831h\,/\,-\,1       & HRI & Fabbiano  & 162066 \\
33 & 600220h                & HRI & Sarazin   &  8538 \\
34 & 600940h                & HRI & Hanlan    & 63877 \\
35 & 202069h                & HRI & Rosner    & 30193 \\
36 & 202070h                & HRI & Rosner    & 29283 \\
37 & 201413h\,-\,1          & HRI & Harnden   & 24202 \\
38 & 202068h                & HRI & Rosner    & 28896 \\
39 & 201414h\,-\,1          & HRI & Harnden   & 34160 \\
40 & 202515h                & HRI &Neuh\"auser& 42286 \\
41 & 202156h                & HRI &Neuh\"auser& 7698 \\
42 & 201090h                & HRI & Damiani   & 5150 \\
43 & 201835h                & HRI & Montmerle & 51710 \\
44 & 201618h\,-\,1\,/\,-\,2 & HRI & Zinnecker & 5566 \\
45 & 201834h\,/\,-\,1\,/\,-\,2 & HRI & Montmerle & 77869 \\
46 & 201709h - 201714h      & HRI & Damiani   & 28031 \\ 
47 & 201055h                & HRI & Walter    & 3650 \\
48 & 201395h\,/\,-\,1       & HRI & Walter    & 19641 \\ \hline

\end{tabular}

\end{table}


The low-mass object V410~x-ray~3 with spectral type M6-7,
a mass of only $\le 0.15~M_{\odot}$, and an age of just $\sim 1$ Myr 
(Luhman et al. 1998) is clearly detected in X-rays (Strom \& Strom 1994).
The optical and infrared data by Luhman et al. (1998) have
recently been re-evaluated by Brice\~no et al. (1998), who classify
this object as brown dwarf candidate with a mass of $0.03$ to $0.08~M_{\odot}$ 
and an age of $\le 0.9$ to $1.5$ Myr, depending on which evolutionary
tracks and isochrones are used. The classification of this X-ray source
as brown dwarf candidate shows that an object near the stellar burning
limit can emit X-rays.

Recently, CRN99 have performed an $H \alpha$ objective prism survey 
as well as optical and infrared photometry of the Cha I dark cloud, 
a site of on-going low- and intermediate-mass star formation. 
They found three low-mass objects
with infrared excess and six additional low-mass late-type objects 
with $H \alpha$ emission, all of which are near, and some possibly below 
the hydrogen burning limit. NC98 did follow-up spectroscopy with high
S/N of the lowest-mass object, $Cha~H \alpha$~{\em 1}, and estimated
its spectral type to be M7.5-M8. Using IJHK photometry (CRN99)
and a distance of 160~pc (ie., the mean of the HIPPARCOS distances
of Cha I T~Tauri stars, Wichmann et al. 1999), its location in the H-R 
diagram is below the hydrogen burning limit; 
a first comparison with evolutionary tracks and isochrones preliminary 
yielded an age of $\sim 1$ Myr and a mass of $\sim 0.04~M_{\odot}$ (NC98),
recently revised by CRN99 to an age of $0.4 \pm 0.1$ Myrs
and a mass of $0.03 \pm 0.01~M_{\odot}$.

NC98 found that this 
object is clearly detected in a 36~ks ROSAT pointed observation
of the Cha I dark cloud. Two Cha I brown dwarf candidates are marginally 
detected, one is not resolved from a nearby star, and the other 
candidates, including all three with infrared excess, are undetected.
Thus, there is evidence that young brown dwarfs show X-ray 
emission, and that it may be rewarding to engage in a wider survey.

In addition, two of the brown dwarf candidates in $\rho$ Oph
recently presented by Wilking et al. (1999), namely GY~5 and GY~37,
are listed as tentative counterparts to X-ray sources by 
Casanova et al. (1995), as already mentioned by Wilking et al. (1999).

There are two more motivations for studying 
whether brown dwarfs can emit X-rays, namely:
First, Pravdo et al. (1996) argue from the low X-ray luminosity 
of the 51 Peg system that the companion to 51 Peg is a planet.
If 51 Peg B would be a close stellar companion, they argue, 
then one would expect stronger X-ray emission.
However, they also come to the conclusion that the upper mass
limit of 51 Peg B is $\sim 10~M_{jup}$, so that -- depending on 
the distinction between giant planets and brown dwarfs -- 51 Peg B 
may either be a brown dwarf or a planet. Because it is not at all clear
whether brown dwarfs usually emit X-rays, the conclusion drawn by 
Pravdo et al. (1996) may not be correct. If brown dwarfs can emit 
X-rays, the nature of 51 Peg B would remain unclear.
Secondly, De Paolis et al. (1998) argued that Massive Astrophysical Compact
Halo Objects (MACHOs) are dark clusters of brown dwarfs, which emit
X-rays and can therefore be responsible for some part of the observed
diffuse X-ray background. 

\section { The ROSAT data }

In this study, we include only isolated brown dwarfs,
ie. brown dwarfs which are not companions to normal stars.
Hence, we include PPl~15, 
a double-lined spectroscopic binary consisting of two brown dwarfs 
(Basri \& Mart\'\i n 1998b), and also the visual binary CFHT-Pl~18, 
where both companions are brown dwarfs (Mart\'\i n et al. 1998c).
Brown dwarf companions to normal stars are all close to their 
primary stars (a few arc seconds or less);
and because all these primaries have late spectral types,
they are all X-ray bright stars, so that
possible X-ray emission from their brown dwarf companions cannot be 
resolved with the ROSAT Positional Sensitive Proportional Counter (PSPC) 
or the High Resolution Imager (HRI), although the latter offers a spatial
resolution of $\sim 4$ arc seconds. For details on ROSAT and its
instruments, the PSPC and the HRI, we refer to Tr\"umper (1982),
Pfeffermann et al. (1988), and David et al. (1996), respectively.
Even the recently detected brown dwarf companion $16$ arc seconds south-west 
of G~196-3 (Rebolo et al. 1998), which has been observed with the PSPC 
in the ROSAT All-Sky Survey (RASS), is not resolvable as possible faint
X-ray source from its primary, an X-ray bright star listed in the 
ROSAT Bright Source Catalog (Voges et al. 1996).

In Tables 1 and 2, we list 26 previously published brown dwarfs and 57 
brown dwarf candidates, respectively, sorted by right ascension in each group,
with designation, bolometric luminosity, distance, and references. 
The Cha I objects are excluded here,
because they were studied in CRN99 and NC98.
In the last line of Table 2, we also list BRI~0021$-$0214, 
which is not a brown dwarf, but an M9.5 star included in our study, 
because it rotates particularly fast (Basri \& Marcy 1995),
which might induce strong X-ray emission due to a rotation
driven dynamo effect. We have recently obtained a new, 64~ks
ROSAT HRI observation centered on BRI~0021$-$0214.
Bolometric luminosities $L_{bol}$ as given in Table 2 for objects in CrA 
as well as for CRBR~28 and 33 in $\rho$ Oph
are calculated as in Comer\'on et al. (1998a), scaled to 
130~pc (160~pc) distance for CrA ($\rho$ Oph) and include a correction for 
foreground extinction. The $L_{bol}$ values for the Hawkins et al. objects 
(the last two field brown dwarf candidate entries in Table 2;
their third object, D12, is omitted here, because a spectrum taken by 
Mart\'\i n \& Basri, in preparation, shows that it is not a brown dwarf) 
are calculated from their J magnitudes in the near-IR, 
assuming a bolometric correction of $B.C. = -2$ at J
(see Lawson et al. 1996 and references therein), based on the 
the extremely late spectral types indicated by their infrared colors.
At J, the correction is not affected by any important absorption features,
even for temperatures well below 2000 K, as observed in Gl~229~B,
which ensures a smooth behavior of B.C. as a function of temperature
(see Allard et al. 1997, Allard \& Hauschildt 1995).
Even for a spectral type later than M9, 
it is not likely to be off by more than 0.1 mag. 
$L_{bol}$ for Pleiades objects with {\em MHO} designation as well as 
for Roque 5, 12, and 25 are estimated as in Zapatero-Osorio et al. (1997b).

In Table 3, we list all the ROSAT PSPC (first group) and HRI (second group) 
pointed observations analyzed here (sorted by right ascension in each group); 
we assign a running number (to be used in Tables 4 to 6)
and list also the official pointing ID, the instrument used, 
the Principle Investigator (PI), and the nominal exposure time.


\begin{table*}

\begin{tabular}{lcccc|lrlcc} 
\multicolumn{10}{c}{\bf Table 4: X-ray upper limits for undetected bona-fide brown dwarfs (except Cha I).} \\ \hline
Object & Spec & $W_{\lambda}(H \alpha)$ & $v \cdot \sin i$ & ref. & FOV & exp. & X-ray & $\log L_{X}$ & $\log$ \\
designation & type & [\AA ] & $[km~s^{-1}]$ & & no. & [ks] & counts & $[erg~s^{-1}]$ & $L_{X}/L_{bol}$\\ \hline 

Roque 4     & M9 &$\le 5$& & 1 & PSPC 3-6,8,11     & 76.2 & \multicolumn{3}{l}{unresolved (a)}     \\ \hline
MHObd3      & M8 & 5.9   & & 2 & RASS              & 0.38 & $\le 0.3$  & $\le 28.18$ & $\le -2.38$ \\ \hline
Roque 5     & M9 &$\le 8$& & 3 & PSPC 2-6,8,9      & 68.6 & $\le 2.9$  & $\le 26.90$ & $\le -3.24$ \\ \hline
Roque 13    &M7.5&10.5&    & 1 & PSPC 2-6,8,9,11   & 76.6 &$\le 17.5$  & $\le 27.64$ & $\le -2.95$ \\ \hline
Roque 11    & M8 & 5.8&    & 1 & PSPC 2-6,8-11     & 103.9& $\le 2.9$  & $\le 26.72$ & $\le -3.72$ \\ \hline
Teide 1     & M8 & 6  &    & 4 & PSPC 2-6,9-11     & 98.6 & $\le 5.0$  & $\le 26.98$ & $\le -3.43$ \\ \hline
Roque 17    &M6.5& 15 &    & 1 & RASS              & 0.36 & $\le 0.3$  & $\le 28.20$ & $\le -2.56$ \\ \hline
Roque 16    & M7 & 5  &    & 2 & PSPC 2,3,5,6,8-11 & 94.1 & $\le 3.8$  & $\le 26.89$ & $\le -3.75$ \\ \hline
PPl 15      &M6.5& 12:&    & 5 & PSPC 2,3,5-10     & 73.8 & $\le 3.0$  & $\le 26.89$ & $\le -3.90$ \\ 
            &    &    &    &   & HRI 38,39         & 59.4 &$\le 42.1$  & $\le 28.61$ & $\le -2.18$ \\ \hline
Roque 12    &M7.5& 19.7&   & 3 & PSPC 2,5,6,8-11   & 68.3 & $\le 3.0$  & $\le 26.92$ & $\le -3.53$ \\ \hline
Roque 25    &L (c)&$\le 5$&& 3 & PSPC 2,7          & 0.80 & $\le 3.6$  & $\le 28.93$ & $\le -0.76$ \\ \hline
Calar 3     & M8 & 10 &    & 6 & PSPC 2,3,8,10     & 16.3 & $\le 4.7$  & $\le 27.74$ & $\le -2.74$ \\ 
            &    &    &    &   & HRI 40            & 42.7 & $\le 0.5$  & $\le 26.82$ & $\le -3.66$ \\ \hline
Teide 2     &M6-7& 7  & 13 &2,6& PSPC 10           & 18.0 & \multicolumn{3}{l}{unresolved (a)}     \\ \hline
CFHT-PL-18  & M8 &    &    & 7 & PSPC 7            & 0.37 & $\le 2.7$  & $\le 29.14$ & $\le -0.82$ \\ \hline
CFHT-PL-12  & M8 & 17 &    & 2 & PSPC 7            & 0.27 & $\le 7.1$  & $\le 29.70$ & $\le -1.07$ \\
            &    &    &    &   & RASS (b)          & 0.37 & $\le 4.0$  & $\le 29.31$ & $\le -1.46$ \\ \hline
CFHT-PL-15  & M7 & 7  &    & 2 & RASS (b)          & 0.36 & $\le 0.3$  & $\le 28.20$ & $\le -2.23$ \\ \hline
LP~944-20   & M9 & 1  &    & 8 & PSPC 20-23        & 66.7 &$\le 84.4$  & $\le 25.58$ & $\le -4.16$ \\
            &    &    &    &   & HRI 31-34         &220.8 &$\le 91.6$  & $\le 25.57$ & $\le -4.17$ \\ \hline
DenisJ1228$-$1547 & L (c) &$\le 1$& 20 & 9,10 & RASS (b)& 0.32&$\le 3.1$&$\le 27.29$ & $\le -2.00$ \\ \hline
Kelu 1      & L (c)& yes & & 9,11 & RASS (b)       & 0.16 & $\le 2.8$  & (d)         &             \\ \hline
CRBR 14     & M7.5 &     & & 12  & PSPC 26         & 30.5 & $\le 6.6$  & $\le 28.13$ & $\le -3.88$ \\
            &      &     & &     & HRI 43-46       & 148.0& $\le 3.9$  & $\le 27.69$ & $\le -4.32$ \\ \hline
GY 10       & M8.5 &     & & 12  & PSPC 26 (e)     & 30.7 & $\le 7.5$  & $\le 28.18$ & $\le -3.97$ \\
            &      &     & &     & HRI 43-46       & 148.0& $\le 3.6$  & $\le 27.65$ & $\le -4.50$ \\ \hline
GY 11 (f)   & M6.5 &     & & 12  & PSPC 26         & 30.8 &$\le 12.6$  & $\le 28.54$ & $\le -2.36$ \\
            &      &     & &     & HRI 43-46       & 148.0& $\le 3.2$  & $\le 27.60$ & $\le -3.30$ \\ \hline
GY 64       & M8   &     & & 12  & PSPC 26         & 31.4 & $\le 6.2$  & $\le 28.09$ & $\le -3.43$ \\
            &      &     & &     & HRI 43-46       & 148.0&$\le 13.0$  & $\le 28.21$ & $\le -3.31$ \\ \hline
GY 141      & M8.5 & 60  & & 12  & PSPC 26 (g)     & 33.0 & $\le 8.7$  & $\le 27.91$ & $\le -3.26$ \\ 
            &      &     & &     & HRI 43-46       & 148.0&$\le 11.5$  & $\le 27.86$ & $\le -3.31$ \\ \hline
GY 310      & M8.5 &     & & 13  & PSPC 26 (h)     & 28.2 &$\le 49.5$  & $\le 29.04$ & $\le -3.29$ \\
            &      &     & &     & HRI 45 (h)      & 76.5 & $\le 7.9$  & $\le 28.28$ & $\le -4.05$ \\ \hline

\end{tabular}

\vspace{-.3cm}

{\small
References: (1) Zapatero-Osorio et al. 1997b, (2) Stauffer et al. 1998a,
(3) Mart\'\i n et al. 1998b, (3) Rebolo et al. 1995, (5) Mart\'\i n et al. 1996,
(6) Mart\'\i n et al. 1998a, (7) Mart\'\i n et al. 1998c, (8) Tinney 1998, 
(9) Mart\'\i n et al. 1997, (10) Tinney et al. 1997
(11) Ruiz et al. 1997, (12) Wilking et al. (1999), (13) Luhman et al. 1997. \\
Remarks: (a) Located in the wing of a bright X-ray source, hence a very large upper limit.
(b) RASS data are listed, if the object is not located in any pointed observation, 
which is significantly deeper than the RASS exposure. 
(c) New spectral type L for low-luminosity objects with spectral type later than M 
as suggested by Mart\'\i n et al. (1997) and Kirkpatrick et al. (1998).
(d) Distance as yet unknown. 
(e) Listed as possible, X-ray detected $\rho$ Oph cloud member in Casanova et al. (1995), 
as also mentioned in Wilking et al. (1999), but located more than one arc minute
away from the X-ray source, so that the identification is dubious; 
we cannot confirm the X-ray detection.
(f) Luminosities are uncertain, because this object seems to be variable in
the infrared (see Comer\'on et al. 1993, 1998a, Wilking et al. 1999).
(g) Slightly lower upper limit reported earlier by Luhman et al. (1997), 
obtained from the $1~\sigma$ noise level in the hard band ROSAT map published 
by Casanova et al. (1995). 
(h) Upper limits are large, because GY~310 is located in the wing of the bright 
X-ray source ROXR1~50 (Casanova et al. 1995), also detected by the 
{\em Einstein Observatory} as source ROX~20 (Montmerle et al. 1983),
identified with the classical T~Tauri star GY 314. 
}

\end{table*}


We reduced all these pointings with 
the Extended Scientific Analysis Software (EXSAS, Zimmermann et al. 1994) 
version 98APR running under ESO-MIDAS version 97NOV.
Whenever an object was found to be located in the field-of-view (FOV) of more than one 
PSPC or HRI pointing, we merged the data sets (separately for the two instruments)
to gain sensitivity.
We performed standard local and map source detection in five different bands:
soft ($0.1$ to $0.4~keV$), hard 1 ($0.5$ to $0.9~keV$), hard 2 ($0.9$ to $2.0~keV$),
hard ($0.5$ to $2.0~keV$), and broad ($0.1$ to $2.0~keV$). 
After merging the source lists, each source was again tested in the above mentioned
five bands by a maximum likelihood 
source detection algorithm.
Eg., a maximum likelihood of existence 
$ML = 14.3$ (or $5.9$) corresponds to 5 (or 3) Gaussian $\sigma$ detections.
We have also reduced the RASS observations in a similar manner for all objects.
The brown dwarf candidates V410 Anon 13 and V410 x-ray 3 are located near the 
bright X-ray source V410 Tau, in the center of PSPC pointing no. 13, but near 
the edge of pointing no. 12, 14, 15, and 16, so that the 
FWHM of the source V410 Tau would be too large in the merged data set;
hence, for these two objects, we use only pointing no. 13.


\begin{table*}

\begin{tabular}{llccrrrccccc} 
\multicolumn{12}{c}{ {\bf Table 5: X-ray detected bona-fide brown dwarfs (BD) and 
candidate brown dwarfs (BDC) (except Cha I). }} \\ \hline
Designation & \hspace{-.4cm} FOV & \multicolumn{2}{c}{X-ray position} & off- & \hspace{-.3cm} ML & \hspace{-.3cm} exp. & \multicolumn{2}{c}{hardness ratios} & \hspace{-.3cm} X-ray & \hspace{-.3cm} $\log L_{X}$ & \hspace{-.3cm} $\log$ \\ 
(Spec type) & \hspace{-.4cm} no. (1) & \hspace{-.2cm} $\alpha _{2000}$ & \hspace{-.3cm} $\delta _{2000}$& set & \hspace{-.3cm} (2) & \hspace{-.3cm} [ks] & $HR~1$ & $HR~2$ & \hspace{-.3cm} counts & \hspace{-.3cm} $[erg~s^{-1}]$ & \hspace{-.3cm} $L_{X}/L_{bol}$ \\ \hline

GY 202 & \hspace{-.4cm} P26 (3) & \hspace{-.3cm} 16:27:04.6 & \hspace{-.3cm} -24:28:36.7 & 19" & \hspace{-.3cm} 8.5 & \hspace{-.3cm} 31.6 & $\ge -0.17$ & $0.86 \pm 0.64$ & \hspace{-.3cm} $7.9 \pm 5.9$ & \hspace{-.3cm} $28.19$ & \hspace{-.3cm} $-3.50$ \\

(M7) BD & \hspace{-.4cm} H43-46 & \multicolumn{4}{c}{ not detected } & \hspace{-.3cm} $148.0$ & & & \hspace{-.3cm} $\le 13.6$ & \hspace{-.3cm} $\le 28.23$ & \hspace{-.3cm} $\le -3.46$ \\ \hline

V410 x-ray 3 & \hspace{-.4cm} P13 (4) & \hspace{-.3cm} 04:18:08.4 & \hspace{-.3cm} 28:26:00.3 & 7" & \hspace{-.3cm} 77.1 & \hspace{-.3cm} 29.8 & $0.90 \pm 0.23$ & $0.00 \pm 0.16$ & \hspace{-.3cm} $60.6 \pm 9.7$ & \hspace{-.3cm} $28.68$ & \hspace{-.3cm} $-3.70$ \\ 

(M6-6.5) BDC & \hspace{-.4cm} H40 & \multicolumn{4}{c}{ not detected } & \hspace{-.3cm} $6.8$ & & & \hspace{-.3cm} $\le 9.4$ & \hspace{-.3cm} $\le 28.99$ & \hspace{-.3cm} $\le -3.39$ \\ \hline

V410 Anon 13 & \hspace{-.4cm} P13 & \hspace{-.3cm} 04:18:18.1 & \hspace{-.3cm} 28:28:40.4 & 12" & \hspace{-.3cm} 11.4 & \hspace{-.3cm} 29.9 & $\ge 0.05$ & $0.72 \pm 0.39$ & \hspace{-.3cm} $11.5 \pm 6.7$ & \hspace{-.3cm} $28.26$ & \hspace{-.3cm} $-3.55$ \\

(M6-6.5) BDC & \hspace{-.4cm} H40 & \multicolumn{4}{c}{ not detected } & \hspace{-.3cm} $6.9$ & & & \hspace{-.3cm} $\le 2.1$ & \hspace{-.3cm} $\le 28.63$ & \hspace{-.3cm} $\le -3.18$ \\ \hline

Tau MHO-4 & \hspace{-.4cm} P17-19 & \hspace{-.3cm} 04:31:24.2 & \hspace{-.3cm} 18:00:22.3 & 3" & \hspace{-.3cm} 94.0 & \hspace{-.3cm} 26.3 & $\ge 0.68$ & \hspace{-.3cm} $-0.12 \pm 0.14$ & \hspace{-.3cm} $82.3 \pm 13.0$ & \hspace{-.3cm} $28.87$ & \hspace{-.3cm} $-3.35$ \\ 

(M6-6.5) BDC & \hspace{-.4cm} H41 & \multicolumn{4}{c}{ not detected } & \hspace{-.3cm} $4.7$ & & & \hspace{-.3cm} $\le 8.9$ & \hspace{-.3cm} $\le 29.13$ & \hspace{-.3cm} $\le -3.09$ \\ \hline

Tau MHO-5 & \hspace{-.4cm} P17-19 & \hspace{-.3cm} 04:32:15.2 & \hspace{-.3cm} 18:12:47.4 & 12" & \hspace{-.3cm} 7.0 & \hspace{-.3cm} 22.9 & $\ge 0.29$ &  $0.15 \pm 0.39$ & \hspace{-.3cm} $14.4 \pm  5.7$ & \hspace{-.3cm} $28.17$ & \hspace{-.3cm} $-3.89$ \\ 

(M6-6.5) BDC & \hspace{-.4cm} H41 & \multicolumn{4}{c}{ not detected } & \hspace{-.3cm} $4.9$ & & & \hspace{-.3cm} $\le 2.6$ & \hspace{-.3cm} $\le 28.57$ & \hspace{-.3cm} $\le -3.49$ \\ \hline

\end{tabular}

\vspace{-.3cm}

{\small
Remarks: 
(1) P for PSPC, H for HRI.
(2) ML is maximum likelihood of existence (see text).
(3) Not listed in Casanova et al. (1993) because of the low S/N ratio,
but with $ML = 8.5$, ie. $4~\sigma$ significance, the source probably is not spurious.
(4) Different data reported by Strom \& Strom (1994) and quoted in Luhman et al. (1998), 
but Preibisch \& Zinnecker (1994) noted inconsistencies 
in the data reduction of Strom \& Strom (1994).
}

\end{table*}


For all sources detected with $ML \ge 7$ (ie. $\ge 3.5~\sigma$),
we have checked whether any of the brown dwarfs or candidates is located within
one arc minute for RASS data, 30 arc seconds for PSPC data, or ten arc seconds
for HRI data, according to their respective positional precision.
One out of 26 bona-fide brown dwarfs is detected in X-rays,
and four out of 57 brown dwarf candidates are detected.
Hence, for all the undetected objects, we have calculated X-ray upper limits 
at the source positions (see Neuh\"auser et al. 1995).

In Tables 4 and 6, we list the X-ray upper limits for bona-fide brown dwarfs
and candidates, respectively, with object designations, running numbers
(from Table 3) of the pointings in which the objects were observed (and/or RASS),
effective exposure times, upper limits to the background subtracted broad band counts,
upper limit X-ray luminosities, and the upper limit X-ray to bolometric luminosity ratios.
In Table 4, we also list the spectral types, $H \alpha$ equivalent widths (positive if in
emission), and projected rotational velocities for the bona-fide brown dwarfs,
as these parameters might be related to X-ray activity.

In Table 5, we list the X-ray data of the detected objects, with object 
designations, spectral types, ROSAT pointing numbers, X-ray position (J2000.0),
offsets between X-ray and optical position, effective exposure times, 
X-ray hardness ratios, background subtracted broad band counts,
X-ray luminosities, and X-ray to bolometric luminosity ratios.
Hardness ratios are X-ray colors defined as follows:
If $Z_{s,m,h}$ are the count rates in the bands soft ($0.1$ to $0.4~keV$), 
medium ($0.5$ to $0.9~keV$), and hard ($0.9$ to $2.0~keV$), respectively, 
then
\begin{displaymath}
HR~1~=~\frac{ Z_{h} + Z_{m} - Z_{s} } { Z_{h} + Z_{m} + Z_{s} }
\quad \mbox{and} \quad
HR~2~=~\frac{ Z_{h} - Z_{m} } { Z_{h} + Z_{m} }
\end{displaymath}
Ie., hardness ratios range from $-1$ to $+1$.
If no counts are detected, e.g, in the soft band, then $HR~1=~1$,
but one can estimate a lower limit to $HR~1$ by using the upper limit
to the soft band count rate $Z_{s}$ in the formula above.

To convert X-ray count rates to fluxes, one must divide the count rates by the appropriate
energy conversion factor, depending on X-ray spectrum and instrument response. 
We assume that the X-ray emission of brown dwarfs is consistent with a
one-temperature Raymond-Smith spectrum (Raymond \& Smith 1977), a thermal spectrum
from a hot, optically thin plasma of solar abundance. 
We use $1~keV$, ie., $\sim 10^{7}~K$, as temperature of the X-ray emitting plasma, 
which is typical for late-type stars (Neuh\"auser et al. 1995, CRN99) 
and the X-ray detected brown dwarfs in Cha I (NC98).
Because foreground absorption is negligible for most of our objects,
we use an energy conversion factor of $10 ^{11}~cts~cm ^{2}~erg ^{-1}$
for PSPC observations (note that the RASS has been obtained with the PSPC).
For the bona-fide and candidate brown dwarfs in $\rho$ Oph, however, 
absorption is not negligible, but typically a few mag (Comer\'on et al. 1998a, 
Wilking et al. 1999), so that we use an appropriately smaller energy 
conversion factor, namely $0.5 \cdot 10 ^{11}~cts~cm ^{2}~erg ^{-1}$; 
except for GY~141, where absorption is negligible, 
namely $A_{V} \simeq 0$ mag (Comer\'on et al. 1998a).
For the Taurus brown dwarf candidates, absorption is very low for all objects,
but V410 Anon 13 (Strom \& Strom 1994, Brice\~no et al. 1998, Luhman et al. 1998).
For hard Raymond-Smith spectra, as assumed here, the energy conversion factor
for HRI observations is roughly three times smaller than for PSPC data.


\begin{table}

\begin{tabular}{llrlll} 
\multicolumn{6}{c}{\bf Table 6: Upper limits for brown dwarf candidates. } \\ \hline
Object & \hspace{-.3cm} FOV  & \hspace{-.3cm} exp. &\hspace{-.3cm} X-ray & \hspace{-.3cm} $\log L_{X}$ &\hspace{-.4cm} $\log$ \\
designation & \hspace{-.3cm} no. & \hspace{-.3cm} [ks] &\hspace{-.3cm} counts &\hspace{-.3cm} $[erg~s^{-1}]$ &\hspace{-.4cm} $L_{X}/L_{bol}$ \\ \hline

PC~0025$+$0447   & \hspace{-.3cm} 1 (1)  &\hspace{-.3cm} 8.7  &\hspace{-.3cm} $\le 3.5$  &\hspace{-.3cm} $\le 27.24$ &\hspace{-.4cm} $\le -2.61$ \\
296~A            & \hspace{-.3cm} RASS   &\hspace{-.3cm} 0.41 &\hspace{-.3cm} $\le 2.8$  &\hspace{-.3cm} $\le 28.22$ &\hspace{-.4cm} $\le -2.49$ \\
J0205$-$1159     & \hspace{-.3cm} RASS   &\hspace{-.3cm} 0.36 &\hspace{-.3cm} $\le 2.6$  &\hspace{-.3cm} $\le 27.34$ &\hspace{-.4cm} $\le -2.25$ \\
AP 270           & \hspace{-.3cm} 30     &\hspace{-.3cm} 16.2 &\hspace{-.3cm} $\le 5.1$  &\hspace{-.3cm} $\le 28.52$ &\hspace{-.4cm} $\le -2.40$ \\
CFHT-PL-8        & \hspace{-.3cm} 4,11   &\hspace{-.3cm} 22.9 &\hspace{-.3cm} $\le 16.9$ &\hspace{-.3cm} $\le 28.15$ &\hspace{-.4cm} $\le -2.57$ \\ 
CFHT-PL-8        & \hspace{-.3cm} 35     &\hspace{-.3cm} 26.7 &\hspace{-.3cm} $\le 2.7$  &\hspace{-.3cm} $\le 27.76$ &\hspace{-.4cm} $\le -2.96$ \\ 
CFHT-PL-17       & \hspace{-.3cm} 4,5,11 &\hspace{-.3cm} 19.3 &\hspace{-.3cm} $\le 5.8$  &\hspace{-.3cm} $\le 27.76$ &\hspace{-.4cm} $\le -2.66$ \\
CFHT-PL-17       & \hspace{-.3cm} 35     &\hspace{-.3cm} 27.1 &\hspace{-.3cm} $\le 19.2$ &\hspace{-.3cm} $\le 28.61$ &\hspace{-.4cm} $\le -1.81$ \\ 
Roque 7        & \hspace{-.3cm} 3-6,8,11 &\hspace{-.3cm} 54.6 &\hspace{-.3cm} $\le 4.6$  &\hspace{-.3cm} $\le 27.20$ &\hspace{-.4cm} $\le -3.08$ \\  
CFHT-PL-20       & \hspace{-.3cm} 11     &\hspace{-.3cm} 18.7 &\hspace{-.3cm} $\le 5.8$  &\hspace{-.3cm} $\le 27.77$ &\hspace{-.4cm} $\le -2.58$ \\ 
CFHT-PL-20       & \hspace{-.3cm} 36     &\hspace{-.3cm} 26.1 &\hspace{-.3cm} $\le 21.1$ &\hspace{-.3cm} $\le 28.66$ &\hspace{-.4cm} $\le -1.69$ \\ 
CFHT-PL-16       & \hspace{-.3cm} 4,11   &\hspace{-.3cm} 19.2 &\hspace{-.3cm} $\le 5.3$  &\hspace{-.3cm} $\le 27.72$ &\hspace{-.4cm} $\le -2.86$ \\ 
CFHT-PL-16       & \hspace{-.3cm} 36     &\hspace{-.3cm} 25.0 &\hspace{-.3cm} $\le 18.9$ &\hspace{-.3cm} $\le 28.63$ &\hspace{-.4cm} $\le -1.95$ \\ 
MHObd4           & \hspace{-.3cm} 4,11   &\hspace{-.3cm} 25.0 &\hspace{-.3cm} $\le 10.7$ &\hspace{-.3cm} $\le 27.91$ &\hspace{-.4cm} $\le -2.63$ \\
MHObd1       & \hspace{-.3cm} 3-5,8,9,11 &\hspace{-.3cm} 83.3 &\hspace{-.3cm} $\le 12.4$ &\hspace{-.3cm} $\le 27.45$ &\hspace{-.4cm} $\le -3.24$ \\
MHObd1           & \hspace{-.3cm} 37     &\hspace{-.3cm} 23.0 &\hspace{-.3cm} $\le 9.1$  &\hspace{-.3cm} $\le 28.36$ &\hspace{-.4cm} $\le -2.33$ \\
CFHT-PL-19       & \hspace{-.3cm} 11     &\hspace{-.3cm} 14.7 &\hspace{-.3cm} $\le 23.9$ &\hspace{-.3cm} $\le 28.49$ &\hspace{-.4cm} $\le -1.94$ \\ 
Roque 15     & \hspace{-.3cm} 2-6,8,9,11 &\hspace{-.3cm} 97.0 &\hspace{-.3cm} $\le 9.8$  &\hspace{-.3cm} $\le 27.28$ &\hspace{-.4cm} $\le -3.45$ \\
Roque 14    & \hspace{-.3cm} 2-6,8,10,11 &\hspace{-.3cm} 103.4&\hspace{-.3cm} $\le 3.1$  &\hspace{-.3cm} $\le 26.76$ &\hspace{-.4cm} $\le -3.83$ \\
Roque 14         & \hspace{-.3cm} 37     &\hspace{-.3cm} 23.0 &\hspace{-.3cm} $\le 8.6$  &\hspace{-.3cm} $\le 28.33$ &\hspace{-.4cm} $\le -2.26$ \\
NPL 37         & \hspace{-.3cm} 2-6,8-11 &\hspace{-.3cm} 103.9&\hspace{-.3cm} $\le 2.9$  &\hspace{-.3cm} $\le 26.72$ &\hspace{-.4cm} $\le -3.65$ \\
MHObd5         & \hspace{-.3cm} RASS     &\hspace{-.3cm} 0.37 &\hspace{-.3cm} $\le 0.3$  &\hspace{-.3cm} $\le 28.19$ &\hspace{-.4cm} $\le -2.50$ \\
NPL 38         & \hspace{-.3cm} 2,3,5-10 &\hspace{-.3cm} 100.5& \multicolumn{3}{l}{unresolved} \\
NPL 38         & \hspace{-.3cm} 37,38    &\hspace{-.3cm} 57.9 &\hspace{-.3cm} $\le 5.1$  &\hspace{-.3cm} $\le 27.70$ &\hspace{-.4cm} $\le -2.65$ \\
PIZ~1          & \hspace{-.3cm} 2,3,8-11 &\hspace{-.3cm} 56.8 &\hspace{-.3cm} $\le 7.0$  &\hspace{-.3cm} $\le 27.37$ &\hspace{-.4cm} $\le -2.83$ \\ 
NPL 36     & \hspace{-.3cm} 2,3,5,6,8-10 &\hspace{-.3cm} 77.7 &\hspace{-.3cm} $\le 3.0$  &\hspace{-.3cm} $\le 26.87$ &\hspace{-.4cm} $\le -3.58$ \\
CFHT-PL-5      & \hspace{-.3cm} 2,3,8,10 &\hspace{-.3cm} 55.0 &\hspace{-.3cm} $\le 2.8$  &\hspace{-.3cm} $\le 26.99$ &\hspace{-.4cm} $\le -3.89$ \\ 
NPL 40     & \hspace{-.3cm} 2,3,6,8,9,11 &\hspace{-.3cm} 82.9 &\hspace{-.3cm} $\le 8.3$  &\hspace{-.3cm} $\le 27.28$ &\hspace{-.4cm} $\le -2.81$ \\
MHObd6         & \hspace{-.3cm} 2,3,8,10 &\hspace{-.3cm} 63.7 &\hspace{-.3cm} $\le 6.0$  &\hspace{-.3cm} $\le 27.25$ &\hspace{-.4cm} $\le -3.49$ \\
CFHT-PL-1      & \hspace{-.3cm} 2,10     &\hspace{-.3cm} 13.7 &\hspace{-.3cm} $\le 4.4$  &\hspace{-.3cm} $\le 27.78$ &\hspace{-.4cm} $\le -3.38$ \\ 
CFHT-PL-1      & \hspace{-.3cm} 40       &\hspace{-.3cm} 40.3 &\hspace{-.3cm} $\le 10.5$ &\hspace{-.3cm} $\le 28.17$ &\hspace{-.4cm} $\le -2.99$ \\ 
CFHT-PL-7      & \hspace{-.3cm} 2,7,10   &\hspace{-.3cm} 17.6 &\hspace{-.3cm} $\le 11.2$ &\hspace{-.3cm} $\le 28.08$ &\hspace{-.4cm} $\le -2.65$ \\ 
CFHT-PL-6      & \hspace{-.3cm} 10       &\hspace{-.3cm} 15.3 &\hspace{-.3cm} $\le 19.1$ &\hspace{-.3cm} $\le 28.37$ &\hspace{-.4cm} $\le -2.66$ \\ 
CFHT-PL-6      & \hspace{-.3cm} 40       &\hspace{-.3cm} 38.0 &\hspace{-.3cm} $\le 4.1$  &\hspace{-.3cm} $\le 27.73$ &\hspace{-.4cm} $\le -3.30$ \\
CFHT-PL-23     & \hspace{-.3cm} 7,10     &\hspace{-.3cm} 12.4 &\hspace{-.3cm} $\le 5.6$  &\hspace{-.3cm} $\le 27.93$ &\hspace{-.4cm} $\le -2.39$ \\ 
CFHT-PL-2      & \hspace{-.3cm} 7,10     &\hspace{-.3cm} 14.2 &\hspace{-.3cm} $\le 2.7$  &\hspace{-.3cm} $\le 27.56$ &\hspace{-.4cm} $\le -3.45$ \\ 
HHJ~22         & \hspace{-.3cm} 7,10     &\hspace{-.3cm} 0.72 &\hspace{-.3cm} $\le 11.5$ &\hspace{-.3cm} $\le 29.48$ &\hspace{-.4cm} $\le -1.50$ \\ 
CFHT-PL-4      & \hspace{-.3cm} 7,10     &\hspace{-.3cm} 0.44 &\hspace{-.3cm} $\le 4.6$  &\hspace{-.3cm} $\le 29.28$ &\hspace{-.4cm} $\le -1.66$ \\
CFHT-PL-26     & \hspace{-.3cm} 7        &\hspace{-.3cm} 0.41 &\hspace{-.3cm} $\le 2.3$  &\hspace{-.3cm} $\le 29.03$ &\hspace{-.4cm} $\le -0.93$ \\
CFHT-PL-25     & \hspace{-.3cm} 7        &\hspace{-.3cm} 0.13 &\hspace{-.3cm} $\le 6.9$  &\hspace{-.3cm} $\le 30.00$ &\hspace{-.4cm} $\le -0.17$ \\
CFHT-PL-25     & \hspace{-.3cm} RASS     &\hspace{-.3cm} 0.38 &\hspace{-.3cm} $\le 0.19$ &\hspace{-.3cm} $\le 27.98$ &\hspace{-.4cm} $\le -1.19$ \\
CFHT-PL-22     & \hspace{-.3cm} RASS     &\hspace{-.3cm} 0.36 &\hspace{-.3cm} $\le 0.14$ &\hspace{-.3cm} $\le 27.87$ &\hspace{-.4cm} $\le -2.33$ \\
RPr 1          & \hspace{-.3cm} 24,25    &\hspace{-.3cm} 18.6 &\hspace{-.3cm} $\le 2.8$  &\hspace{-.3cm} $\le 27.78$ &\hspace{-.4cm} $\le -2.21$\\
GY 5           & \hspace{-.3cm} 26       &\hspace{-.3cm} 30.5 &\hspace{-.3cm} $\le 8.6$  &\hspace{-.3cm} $\le 28.24$ &\hspace{-.4cm} $\le -4.15$\\
GY 5           & \hspace{-.3cm} 43-46    &\hspace{-.3cm} 148.0&\hspace{-.3cm} $\le 7.0$  &\hspace{-.3cm} $\le 27.94$ &\hspace{-.4cm} $\le -4.45$\\
CRBR 28        & \hspace{-.3cm} 26       &\hspace{-.3cm} 31.9 &\hspace{-.3cm} $\le 7.7$  &\hspace{-.3cm} $\le 28.17$ &\hspace{-.4cm} $\le -4.23$\\
CRBR 28        & \hspace{-.3cm} 43-46    &\hspace{-.3cm} 148.0&\hspace{-.3cm} $\le 6.8$  &\hspace{-.3cm} $\le 27.93$&\hspace{-.4cm} $\le -4.46$\\
CRBR 33        & \hspace{-.3cm} 26       &\hspace{-.3cm} 30.6 &\hspace{-.3cm} $\le 12.3$ &\hspace{-.3cm} $\le 28.40$&\hspace{-.4cm} $\le -3.77$\\
CRBR 33        & \hspace{-.3cm} 43-46    &\hspace{-.3cm} 148.0&\hspace{-.3cm} $\le 4.6$  &\hspace{-.3cm} $\le 27.76$ &\hspace{-.4cm} $\le -4.41$\\
GY 31          & \hspace{-.3cm} 26       &\hspace{-.3cm} 30.6 &\hspace{-.3cm} $\le 13.2$ &\hspace{-.3cm} $\le 28.43$ &\hspace{-.4cm} $\le -5.01$\\ 
GY 31          & \hspace{-.3cm} 43-46    &\hspace{-.3cm} 148.0&\hspace{-.3cm} $\le 3.4$  &\hspace{-.3cm} $\le 27.63$ &\hspace{-.4cm} $\le -5.81$\\ \hline

\end{tabular}

\end{table}

\begin{table}

\begin{tabular}{llrlll} 
\multicolumn{6}{c}{{\bf Table 6: Upper limits for BD candidates} (cont.). } \\ \hline
Object & FOV  & \hspace{-.3cm} exp. &\hspace{-.3cm} X-ray & \hspace{-.3cm} $\log L_{X}$ &\hspace{-.4cm} $\log$ \\
designation & no. & \hspace{-.3cm} [ks] &\hspace{-.3cm} counts &\hspace{-.3cm} $[erg~s^{-1}]$ &\hspace{-.4cm} $L_{X}/L_{bol}$ \\ \hline

GY 37& 26 (2)&\hspace{-.3cm} 31.4 &\hspace{-.3cm} $\le 8.2$  &\hspace{-.3cm} $\le 28.21$ &\hspace{-.4cm} $\le -3.76$\\
GY 37&43-46&\hspace{-.3cm} 148.0 &\hspace{-.3cm} $\le 5.8$  &\hspace{-.3cm} $\le 27.86$ &\hspace{-.4cm} $\le -4.01$\\
GY 59&26 &\hspace{-.3cm} 31.2  &\hspace{-.3cm} $\le 6.0$  &\hspace{-.3cm} $\le 28.08$ &\hspace{-.4cm} $\le -4.22$\\
GY 59&43-46&\hspace{-.3cm} 148.0 &\hspace{-.3cm} $\le 6.8$  &\hspace{-.3cm} $\le 27.93$ &\hspace{-.4cm} $\le -4.37$\\
GY 84&26 &\hspace{-.3cm} 30.6  &\hspace{-.3cm} $\le 6.2$  &\hspace{-.3cm} $\le 28.10$ &\hspace{-.4cm} $\le -4.50$ \\
GY 84&43-46&\hspace{-.3cm} 148.0 &\hspace{-.3cm} $\le 11.1$ &\hspace{-.3cm} $\le 28.14$ &\hspace{-.4cm} $\le -4.46$\\
GY 107&26 &\hspace{-.3cm} 31.2  &\hspace{-.3cm} $\le 6.2$  &\hspace{-.3cm} $\le 28.09$ &\hspace{-.4cm} $\le -4.63$ \\
GY 107&43-46&\hspace{-.3cm} 148.0 &\hspace{-.3cm} $\le 15.6$ &\hspace{-.3cm} $\le 28.29$ &\hspace{-.4cm} $\le -4.43$\\
GY 163&26 &\hspace{-.3cm} 30.9  &\hspace{-.3cm} $\le 4.9$  &\hspace{-.3cm} $\le 27.99$ &\hspace{-.4cm} $\le -4.85$ \\
GY 163&43-46&\hspace{-.3cm} 148.0 &\hspace{-.3cm} $\le 33.8$ &\hspace{-.3cm} $\le 28.63$ &\hspace{-.4cm} $\le -4.21$\\
GY 326&26 &\hspace{-.3cm} 28.1  &\hspace{-.3cm} $\le 27.1$ &\hspace{-.3cm} $\le 28.78$ &\hspace{-.4cm} $\le -4.11$ \\
GY 326&45 &\hspace{-.3cm} 76.3  &\hspace{-.3cm} $\le 5.5$  &\hspace{-.3cm} $\le 28.13$ &\hspace{-.4cm} $\le -4.76$ \\
B185815.3$-$370435 & 27,28 &\hspace{-.3cm} 11.7 & \multicolumn{3}{l}{unresolved} \\
B185815.3$-$370435 & 47,48 &\hspace{-.3cm} 18.0 &\hspace{-.3cm} $\le 3.3$  &\hspace{-.3cm} $\le 28.04$ &\hspace{-.4cm} $\le -3.55$ \\
B185831.1$-$370456 & 27,28 &\hspace{-.3cm} 10.5 &\hspace{-.3cm} $\le 8.5$  &\hspace{-.3cm} $\le 28.21$ &\hspace{-.4cm} $\le -2.78$ \\
B185831.1$-$370456 & 47,48 &\hspace{-.3cm} 18.0 &\hspace{-.3cm} $\le 3.2$  &\hspace{-.3cm} $\le 28.03$ &\hspace{-.4cm} $\le -2.96$ \\
B185839.6$-$365823 & 27,28 &\hspace{-.3cm} 13.1 &\hspace{-.3cm} $\le 8.7$  &\hspace{-.3cm} $\le 28.12$ &\hspace{-.4cm} $\le -2.37$ \\
B185839.6$-$365823 & 47,48 &\hspace{-.3cm} 18.3 &\hspace{-.3cm} $\le 3.5$  &\hspace{-.3cm} $\le 28.06$ &\hspace{-.4cm} $\le -2.43$ \\ 
B185840.4$-$370433 & 27,28 &\hspace{-.3cm} 11.4 &\hspace{-.3cm} $\le 10.9$ &\hspace{-.3cm} $\le 28.28$ &\hspace{-.4cm
} $\le -3.11$ \\
B185840.4$-$370433 & 47,48 &\hspace{-.3cm} 17.7 &\hspace{-.3cm} $\le 5.9$  &\hspace{-.3cm} $\le 28.30$ &\hspace{-.4cm
} $\le -3.09$ \\
B185853.3$-$370328 & 27,28 &\hspace{-.3cm} 11.3 &\hspace{-.3cm} $\le 8.4$  &\hspace{-.3cm} $\le 28.17$ &\hspace{-.4cm} $\le -3.22$ \\
B185853.3$-$370328 & 47,48 &\hspace{-.3cm} 17.9 &\hspace{-.3cm} $\le 6.3$  &\hspace{-.3cm} $\le 28.32$ &\hspace{-.4cm} $\le -3.07$ \\
D04               & RASS  &\hspace{-.3cm} 0.36 &\hspace{-.3cm} $\le 2.0$  &\hspace{-.3cm} $\le 28.13$ &\hspace{-.4cm} $\le -1.23$ \\
D07               & RASS  &\hspace{-.3cm} 0.25 &\hspace{-.3cm} $\le 3.3$  &\hspace{-.3cm} $\le 28.51$ &\hspace{-.4cm} $\le -0.55$ \\
BRI~0021$-$0214   & 29    &\hspace{-.3cm} 63.2 &\hspace{-.3cm} $\le 3.1$  &\hspace{-.3cm} $\le 25.41$ &\hspace{-.4cm} $\le -4.68$ \\ \hline

\end{tabular}

\vspace{-.3cm}

{\small
(1) Similar upper limit reported by Fleming et al. (1993). \\
(2) Listed as possibly X-ray detected $\rho$ Oph cloud member in Casanova et al. (1995), 
as also mentioned in Wilking et al. (1999), but it is located more than one arc minute
away from the X-ray source, so that the identification is dubious;
we cannot confirm the X-ray detection.
}

\end{table}


The detected bona-fide brown dwarf $\rho$ Oph GY~202,
initially suggested as brown dwarf candidate by Comer\'on et al.
(1993, 1998a), was recently confirmed to be a very young ($\le 1$ Myr)
bona-fide brown dwarf by Wilking et al. (1999).
The four X-ray detected brown dwarf candidates are
all located in L1495E, a young star forming cloud in Taurus.
None of the other relatively young and, hence, bright objects, 
located in $\rho$ Oph and CrA, nor any of the Pleiades objects 
were detected, despite the long ROSAT PSPC pointed 
observations in these fields (up to $\sim 100$~ks). 
In the case of the $\rho$ Oph and CrA objects, this could be because they 
have large extinction (except GY~141), hence X-rays in the 
wavelength range available to ROSAT would be highly absorbed. On the
other hand, the Pleiades brown dwarfs are two orders of magnitude older 
than the young Taurus and Cha I objects, so they could simply be too 
faint in X-rays.
It is surprising that $\rho$ Oph GY~202 is detected in spite of
the strong extinction of $A_{V} = 13$ mag (Wilking et al. 1999).
We detected only photons in the hard band (above $\sim 1~keV$),
consistent with strong extinction.

\begin{figure}
\vbox{\psfig{figure=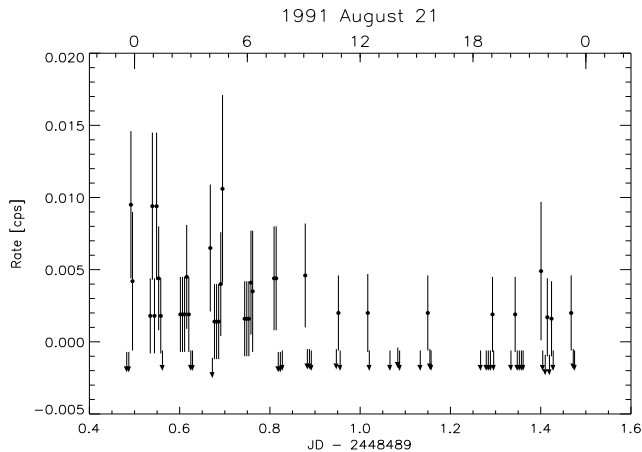,width=9cm,height=6cm}}
\caption{ X-ray light curve of V410 x-ray 3 with count rate in 
counts per second versus the observation time (JD date at the 
bottom axis, UT time on top axis), $400~s$ bins, $1~\sigma$ errors }
\end{figure}

\begin{figure}
\vbox{\psfig{figure=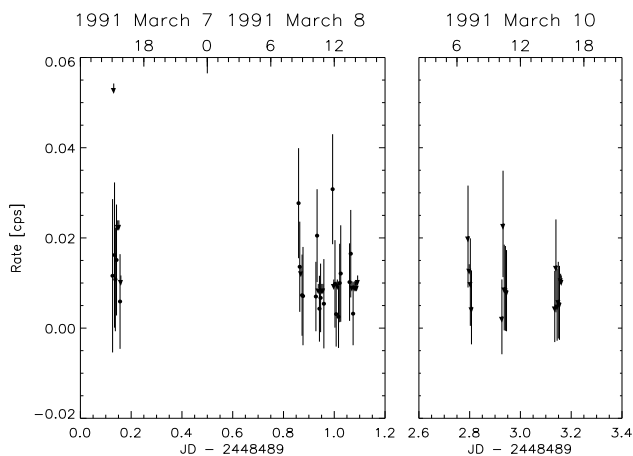,width=9cm,height=6cm}}
\caption{ X-ray light curve of Tau MHO-4 with count rate in counts per
second versus the observation time (JD date at the bottom axis,
UT time on top axis), $400~s$ bins, $1~\sigma$ errors }
\end{figure}

Among the bona-fide brown dwarfs, the lowest upper limit in terms
of $L_{X}$ is found for LP~944-20, namely 
$\log (L_{X}/erg~s^{-1}) \le 25.57$ (ie. $\log (L_{X}/L_{bol}) \le -4.17$)
with HRI observations of 221~ks in total;
in terms of $\log (L_{X}/L_{bol})$, the lowest upper limit
is found for $\rho$ Oph GY~10, namely $\log (L_{X}/L_{bol}) \le -4.50$
with HRI observations of 148~ks in total.
As far as the brown dwarf candidates are concerned, the lowest upper limit 
in terms of $L_{X}$ is found for NPL~37, namely
$\log (L_{X}/erg~s^{-1}) \le 26.72$ (ie. $\log (L_{X}/L_{bol}) \le -3.65$)
with a 104~ks PSPC pointed observations;
in terms of $\log (L_{X}/L_{bol})$, the lowest upper limit
is found for $\rho$ Oph GY~31, namely
$\log (L_{X}/L_{bol}) \le -5.81$
with HRI observations of 148~ks in total.
For BRI~0021$-$0214, we found a lower limit of $\log (L_{X}/erg~s^{-1}) \le 25.41$,
ie. $\log (L_{X}/L_{bol}) \le -4.68$) in a 63~ks HRI observation.

There are now six X-ray detected brown dwarf candidates (four in Taurus, two in Cha I)
and two detected bona-fide brown dwarfs. Out of those eight X-ray detected objects,
only two have a S/N ratio of larger than $5$, namely Tau MHO-4 ($82.3 \pm 13.0$ counts)
and V410 x-ray 3 ($60.6 \pm 9.7$ counts), so that a meaningful timing analysis is 
possible. Tau MHO-4 is detected in three different pointings at slightly different count 
rates, namely $3.2 \pm 1.0$ cts/ks in the PSPC pointing no. 17 (PI Pye, obtained 
in March 1991), $5.9 \pm 1.5$ cts/ks in no. 18 (PI Zinnecker, Sept. 1992), 
and $3.0 \pm 0.7$ cts/ks in no. 19 (PI Burrows, Feb. 1993),
indicating no variability above a $\sim 1~\sigma$ level on a time-scale of months to years.
There is no indication for variability for any of the other detected objects, 
nor in the undetected objects.

We display the X-ray light curves for Tau MHO-4 (PSPC pointing no. 19)
and V410 x-ray 3 (PSPC pointing no. 13) in Fig. 1 and 2, respectively.
No variability 
on a time-scale of hours
can be detected in these light curves.

\section { Comparison with Cha I }

We list the X-ray data of the Cha I bona-fide brown dwarf
($Cha~H \alpha$~{\em 1}) and brown dwarf candidates from CRN99 and 
NC98 in Table 7, all of which are only $\sim 1~Myr$ old.
The X-ray luminosities of the three detected objects is $\sim 10^{28}~erg~s^{-1}$,
the luminosity ratio $\log (L_{X}/L_{bol})$ is in the range $-3.4$ to $-4.3$.
The data found for the four detected brown dwarf candidates in Taurus
are very similar, the average in $\log~L_{X}$ being $28.5$ and in 
$\log (L_{X} / L_{bol})$ being $-3.6$, 
very similar also to $Cha~H \alpha$~{\em 1}~and $\rho$ Oph GY~202.
This seems to be the typical X-ray emission level 
for such young, late-type, low-mass objects.
Note that CRN99 have searched for brown dwarfs only in the center of the Cha I
dark cloud, but nowhere else, so that there should be no strong bias towards
the X-ray brightest low-mass objects in our Cha I sample.


\begin{table}

\begin{tabular}{lllrlcc} 
\multicolumn{7}{c}{\bf Table 7: Cha I low-mass members. } \\ \hline
Desig.                & Spec & exp. & X-ray      & $\log L_{X}$ & $\log$          & ref. \\
{\scriptsize {CRN99}} & type & [ks] & counts     & $[erg~s^{-1}]$    & $L_{X}/L_{bol}$ &      \\ \hline

$H\alpha~1$ & {\scriptsize M7.5-8} & 37.8 & $31.4$ & $28.41$ & $-3.44$ & 1,2 \\
$H\alpha~2$ & M6     & \multicolumn{4}{l}{\scriptsize {not resolved, too close to another star}} & 1,2 \\
$H\alpha~3$ & M6     & 37.6 & $11.9$ & $27.99$ & $-4.31$ & 1,2 \\
$H\alpha~4$ & M6.5   & 34.8 & $\le 22.9$ & $\le 28.31$ & $\le -4.91$ & 1,2 \\
$H\alpha~5$ & M6     & 33.9 & $\le 2.0$ & $\le 27.26$ & $\le -5.24$ & 1,2 \\
$H\alpha~6$ & M6     & 31.8 & $8.2$ & $27.90$ & $-4.09$& 1,2 \\
$IR~1$ & & 30.7 & $\le 4.7$ & $\le 27.68$ & $\le -4.37$ & 2 \\
$IR~2$ & & \multicolumn{4}{l}{\scriptsize {not resolved, too close to another star}} & 2 \\
$IR~3$ & & 34.0 & $\le 5.3$ & $\le 27.68$ & (3) & 2 \\ \hline

\end{tabular}

\vspace{-.3cm}

{\small
Remarks: (1) NC98, (2) CRN99. (3) $L_{bol}$ as yet unknown.
$H\alpha~1$ is a bona-fide brown dwarf (NC98), 
while the other objects are brown dwarf candidates (CRN99).
}

\end{table}


Out of the 26 bona-fide brown dwarfs studied in this paper, 
15 objects have an upper limit
to the X-ray to bolometric luminosity ratio above the value found for 
$Cha~H\alpha$~{\em 1}, so that most of the non-detections found here may be
due to too short X-ray exposures given the low optical/IR luminosities.
The situation is similar for the brown dwarf candidates:
Out of the 57 objects studied, 39 have an upper limit to 
$L_{X}/L_{bol}$ above the value found for $Cha~H\alpha$~{\em 1}.
See also Fig. 3 for a comparison of the upper limits 
found here with some X-ray detections.

Hence, although few of the ROSAT observations investigated here are 
deep enough to reach the $L_{X}/L_{bol}$ values found for low-mass 
members of Cha I, we can conclude that the Pleiades brown dwarfs 
and candidates are X-ray fainter than the Cha I brown dwarfs.
Even at the age of the Pleiades ($\sim 10^{8}$ yrs), 
brown dwarfs seem to be already too old, ie., too faint, 
to be detected in $\sim 100$~ks PSPC observations.

\section {Discussion and conclusion}

The $\log~L_{X}/L_{bol}$ values for the detected objects lie intermediate between 
the brightest and faintest of the nine X-ray detected infrared Class I objects,
which are EC 95 in Serpens with $\log~L_{X}/L_{bol} = -2.0$ (Preibisch 1998)
and TS~2.6 in CrA with $\log~L_{X}/L_{bol} = -4.3$ (Neuh\"auser \& Preibisch 1997).
T~Tauri stars, though, typically have much larger X-ray luminosities,
see Neuh\"auser et al. (1995). X-ray saturation level for T~Tauri stars
and X-ray active other stars is reached at $\log~L_{X}/L_{bol} \simeq -3$.
For most of the objects observed with pointed observations,
the upper limits are below this value.
Also, for all the objects in Chamaeleon and Taurus, the $\log~L_{X}/L_{bol}$
values (or upper limits) lie below $-3$.
For those objects, we can exclude that X-ray emission is saturated.
We cannot exclude, though, that their X-ray emission is below the level 
of the quiet Sun being $\log~L_{X}/L_{bol} \simeq -6$ (see Schmitt 1997).
We find the same for BRI~0021$-$0214. The upper limit shows that its X-ray 
emission cannot be saturated, even though the fast rotation, but it still
lies above the quiet Sun level.
However, in the special case of $\rho$ Oph GY~31, we find an upper limit
of $\log~L_{X}/L_{bol} \simeq -5.81$, ie. very close to the quiet Sun; this 
is more than an order of magnitude lower than for all other objects studied.
The upper limits from both the PSPC and HRI observations are below $-5$.
This is particularly surprising, because this object shows radio emission at 
3.6~cm which may indicate non-thermal gyro-synchrotron emission due to magnetic
field lines, in which hot, X-ray emitting plasma should be trapped. 
This radio emission, though, appears to be highly variable with
$0.5$~mJy in its high state and $\le 0.1$~mJy in its low state
(P. Andr\'e, private communication; incorrectly quoted 
in Wilking et al. 1999 as radio emission at 6~cm).
It may be possible that all ROSAT observations 
of GY~31 took place during low-activity phases.

Canonical stellar evolution theory predicts that stars below 
$\sim 0.3~M_{\odot}$ are fully convective (Drake et al. 1996).
Hence, like low-mass late-type stars, brown dwarfs may emit X-rays; 
and in Cha I, one bona-fide brown dwarf and possibly two brown dwarf 
candidates are detected X-ray sources (NC98).
However, convection alone is not enough to generate magnetic fields 
(and coronal X-ray emission). Some kind of rotation-induced dynamo is needed. 
Differential rotation could drive dynamo activity ($\alpha$-$\omega$ dynamo) 
but convection implies rigid rotation and thus a solar-type dynamo should 
be quenched in completely convective stars (Drake et al. 1996).
However, in the absence of differential rotation, another type of dynamo,
related to the Coriolis force and driven by turbulent convection, 
can take over and cause coronal X-ray activity ($\alpha ^{2}$ dynamo).

If X-ray emission in brown dwarfs is driven by a dynamo, then it may
be that only the fastest rotating brown dwarfs can be detected 
given the sensitivity of current instrumentation.
Projected rotational velocities $v \cdot \sin i$ are known 
for two bona-fide brown dwarfs (see Table 4). 
They rotate relatively fast, but are not observed deeply
with ROSAT (nor are they detected).

While an old star with a very late spectral type, like eg. M9, may in fact display 
weak coronal ($H \alpha$ and X-ray) activity, eg. due to low rotation,
both young low-mass stars and young brown dwarfs with spectral types 
of late M or L can be fully convective and may then show X-ray emission. 
As the brown dwarf ages, fusion of deuterium in its center will stop, 
so that the temperature in its core will decrease.
Consequently, the temperature gradient between center and surface
will decrease 
and, hence, the convection velocity will decrease.
Therefore, old brown dwarfs may not be able to show X-ray emission,
even if they rotate fast,
while old main sequence stars may still be able to produce X-rays.

Perhaps, 
the very low X-ray detection fraction of brown dwarfs is due to 
variability in the sense that brown dwarfs are no or very faint X-ray
emitters for most of the time, but are in an high state for a time
that is short compared to the typical exposure.
However, it is unlikely that we would, by chance, detect X-ray emission 
from several brown dwarfs (or candidates) in Cha I and Taurus, 
while all Pleiades and field brown dwarfs remain undetected.
Also, the X-ray detected objects in Cha I and Taurus do not 
show significant variability.

The X-ray detected brown dwarfs in Cha I may be close binaries with 
magnetic field configurations like in X-ray bright RS~CVn-type binaries.
However, it would be surprising that $Cha~H\alpha$~{\em 1}~(as a 
binary and being the X-ray brightest) is the faintest object in the optical 
among the six $H \alpha$ detected objects in Cha I (CRN99); 
if the other brown dwarf candidates in Cha I were close binaries, 
too, they should be brighter X-ray sources.

The X-ray emission may be linked to accretion or magnetic reconnection
between the brown dwarf and a circumstellar disks, 
but no K-band near-IR excess is seen in the Cha I objects (CRN99).
However, as far as $Cha~H\alpha$~{\em 1}~is concerned,
excess emission is observed at 6.7 and 14.3 $\mu m$ by ISOCAM with a
color index $[14.3/6.7]=0.15$ as in normal Class II objects
(Comer\'on et al. 1998b); the fact that no excess is detected  in the
K-band is probably due to a very low accretion rate.
Also, in the cases of $\rho$ Oph GY~202 as well as Tau MHO-4 and MHO-5, 
some near-IR excess does indicate the presence of circumstellar material 
(Wilking et al. 1999; Brice\~no, unpublished).

\begin{figure}
\vbox{\psfig{figure=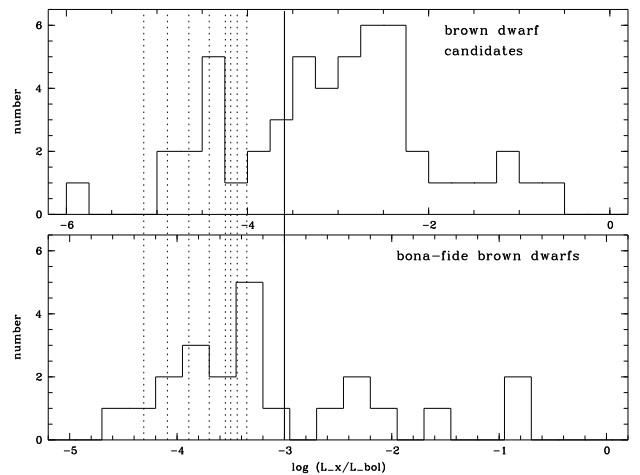,angle=-90,width=10cm,height=7cm}}
\caption{ Histogram of upper limits for $\log~(L_{X} / L_{bol})$ for 
brown dwarf candidates (upper panel) and bona-fide brown dwarfs (lower panel).
Dotted lines indicate the values of detected objects, namely from left to right 
$Cha~H\alpha$~{\em 3, 6}, Tau MHO-5, V410 x-ray 3, V410 Anon 13, $\rho$ Oph GY~202, 
$Cha~H\alpha$~{\em 1}, and Tau MHO-4 (with data from Tables 5 and 7).
The X-ray saturation level is shown as full line. 
For objects with several upper limits available 
in Tables 4 or 6, we plot only the lowest limit }
\end{figure}

$H \alpha$ emission, a proxy for chromospheric activity, might be related to
X-ray emission. Two of the three X-ray detected objects in Cha I show 
strong $H \alpha$ emission with $W_{\lambda}(H \alpha) \simeq 60$\AA~(CRN99),
and also the four detected objects in Taurus show strong $H \alpha$ emission
(Luhman et al. 1998, Brice\~no et al. 1998). However, the brown dwarf 
$\rho$ Oph GY~141 shows similarily strong $H \alpha$ emission (Luhman et al. 1997), 
but no X-ray emission. $H \alpha$ emission is weak in most of the other 
bona-fide brown dwarfs (see Table 4).

X-ray emission of brown dwarfs, if typically present even in brown dwarfs
as old as the Pleiades or older, may be similar to late-type low-mass stars, 
with $\log (L_{X}/L_{bol}) \simeq -4$. This cannot be rejected from our upper limits.
However, it may be possible that only brown dwarfs younger than $\sim 3$ Myrs emit X-rays.
Then, in contrast to normal stars, they get fainter and fainter in $L_{bol}$ 
as they age (Burrows et al. 1995). Hence, their X-ray luminosity should also decrease.
However, because low-mass objects like brown dwarfs probably have only weak
or no winds, they have much longer time-scales for angular momentum loss 
compared to normal stars (Basri \& Marcy 1995, Mart\'\i n \& Zapatero-Osorio 1997).
Hence, if X-ray emission of brown dwarfs is due to a rotationally driven dynamo, 
as in other fully convective low-mass stars, then X-ray luminosities 
should decrease more slowly in brown dwarfs than in normal stars,
so that the non-detection of Pleiades brown dwarfs would be surprising.
We conclude that rotation is not the key parameter in X-ray emission of
brown dwarfs. More important may be the fact that the convection velocity
of brown dwarfs decreases significantly, because they become fainter as they age.
Then, one would also expect $H \alpha$ emission to be weaker in middle-aged and 
old brown dwarfs compared to young brown dwarfs. Indeed, strong $H \alpha$ emission
is observed only in young brown dwarfs, but neither in Pleiades nor field brown dwarfs.
From the X-ray detection of a few $\sim 1$ Myr young brown dwarfs and candidates
and the undetection of the $\sim 100$ Myrs old Pleiades brown dwarfs,
we may conclude that all the nearby field brown dwarfs and candidates,
all undetected in X-rays, are too old and, hence, too faint for being detected 
in the long pointed observations, even though some of them may still rotate very 
fast. Eg., BRI~0021$-$0214 does rotate fast, but is undetected in X-rays 
and $H \alpha$ (Basri \& Marcy 1995). Also, the old field brown dwarf
DenisJ1228$-$1547 rotates fast, but shows weak or no $H \alpha$ emission
(Tinney et al. 1997) as well as faint or no X-ray emission (see table 4).

De Paolis et al. (1998) assumed that brown dwarfs display the same X-ray
luminosity as very late-type stars like vB~8, ie. $\sim 10^{27}~erg~s ^{-1}$.
If MACHOs are clusters of brown dwarfs, then they could contribute for some
part of the observed diffuse X-ray background. Here, we have obtained
$\sim 10^{28}~erg~s ^{-1}$ as typical X-ray emission level for young brown dwarfs.
Although most of our upper limits do not reach below $10^{27}~erg~s ^{-1}$,
we do find significantly lower limits for BRI~0021$-$0214 and some 
bona-fide and candidate brown dwarfs. Hence, even if MACHOs are clusters 
of brown dwarfs, their contribution to the diffuse X-ray background 
may be lower than estimated by De Paolis et al. (1998).

With only a few detections 
(two bona-fide brown dwarfs and six candidates) and 81 upper limits, 
the X-ray luminosity function of brown dwarfs is not well constrained. 
However, there are still thousands of faint X-ray 
sources in deep ROSAT pointed observations, which remain to be identified. 

\begin{appendix}

\section { Appendix: B- and R-band data}

We have cross-correlated the optical/IR positions of all bona-fide
and candidate brown dwarfs studied here with the
United States Naval Observatory (USNO) Precision Measuring Machine 
(PMM) catalog, which lists astrometric positions and 
photographic magnitudes in the blue (emulsion O or J) and red 
(emulsion E or F), all $\pm 0.5$ mag (Monet 1996).
Two of the objects listed above in 
Tables 1 and 2 are found in that catalog.
The field brown dwarf candidate 296~A has $B = 21.0$ mag and 
$R = 17.8$ mag according to the USNO-PMM catalog; 
Thackrah et al. (1997) gave $R_{F} - I_{N} = 2.54 \pm 0.20$ mag 
with $I_{N} = 14.57 \pm 0.15$ mag. For the young Taurus brown dwarf 
candidate MHO-5, the USNO-PMM catalog lists $B = 18.9$ mag and 
$R = 15.7$ mag, whereas Brice\~no et al. (1998) gave $R_{C} = 16.23$ mag.
Some of the Cha I low-mass objects are also listed in the 
USNO-PMM catalog (see note to table 4 in CRN99).
It may well be possible to find more brown dwarf candidates
by cross-correlating the USNO-PMM catalog with other useful
data bases like ROSAT source lists.

\end{appendix}

\acknowledgements{
We would like to thank Tom Fleming, Adam Burrows and Gibor Basri for useful 
discussion about X-ray emission of brown dwarfs. We are also grateful to 
Laurent Cambresy for pointing us to the USNO-PMM catalog.
ROSAT is supported by the German government (BMBF/DLR) and the Max-Planck-Society. 
RN wishes to acknowledge financial support from the DFG Star Formation program. 
ELM was supported by a postdoctoral fellowship of the Spanish MEC.  }

{}

\end{document}